\documentstyle[12pt,aaspp]{article}
\def\@journalname{Astrophysical Journal}
\def\accepted#1{\gdef\@accptdate{#1}} \accepted{\relax}
\def\journalid#1#2{\gdef\@jourvol{#1}\gdef\@jourdate{#2}}
\def\articleid#1#2{\gdef\@startpage{#1}\gdef\@finishpage{#2}}
\def\kms{{\rm\,km\,s^{-1}}}
\def\msun{{\,M_\odot}}

\def\pc{{\rm\,pc}}

\begin{document}

\title{High Altitude Molecular Clouds}
\author{Sangeeta Malhotra}
\affil{Princeton University Observatory, Peyton Hall, Princeton, NJ
08544 \\
 I: san@astro.princeton.edu}

\begin{abstract}
A population of molecular clouds with a significantly greater scale
height than that of Giant Molecular Clouds has been identified by
examining maps of the latitude distribution of the $^{12}CO(1-0)$
emission in the first quadrant of the Galaxy. These clouds are found
by identifying emission more than 2.6 times the scale-height away
from the galactic midplane (centroid of CO emission) at the tangent
points. Since the distance to the tangent points is known, we know the
height and the sizes of these clouds. They are smaller and fainter
than the GMCs and do not seem to be gravitationally bound. These clouds
have properties similar to the high latitude clouds in the solar
neighborhood. Although they lie outside the molecular cloud layer, the
high altitude clouds are well within the HI layer in the Galaxy and
coincide with distinct peaks in the HI distribution. These clouds
represent a galaxy wide population of small molecular clouds having a
larger scale height. They may be clouds in transition between
molecular and atomic phases.

\end{abstract}

\section{Introduction}
Molecular gas is confined to a layer much thinner than the stellar
disk, and if isothermal, is expected to have an equilibrium
distribution which is Gaussian in z (Spitzer 1942). This layer seems
to be Gaussian from the various surveys of molecular gas in the
galactic plane (Sanders et al. 1986, Clemens et al. 1986,Cohen et
al. 1986, Bronfman et al. 1989, Dame et al. 1987). Since the molecular
gas forms a cold, thin layer in the disk, it may be possible to look
for signatures of disequilibrium in departures from Gaussianity of the
distribution.

These departures from a Gaussian distribution in $z$ could be due to
violent events stirring up the gas (for example supernovae) and
subsequent formation of molecular clouds in swept-up shells, or to
short lifetimes of molecular clouds, or to infalling gas etc.
Alternatively, they might reflect the existence of more than one
population of molecular clouds with different (cloud-cloud) velocity
dispersions and hence different scale heights in equilibrium.

We have systematically searched for and found a population of small
clouds lying at high altitudes away from the galactic plane. These
clouds are identified in the $^{12}CO(1 \rightarrow 0)$ survey of
Knapp, Stark \& Wilson (1985). This survey combines a wide latitude
coverage (typically $\pm 2 \deg$) with close sampling in latitude
($\Delta b =2 \arcmin$).

The clouds are identified at the tangent points where we know their
distances and hence can determine their height above the galactic
plane. The vertical distribution of CO emission at tangent points,
i.e. the scale height and mid plane positions have been measured by
Malhotra (1994, hereafter paper I). The data and the model fits to are
described briefly in section 2.1 and 2.2. In section 2.3 the selection
criteria for the identification of the high altitude clouds (HACs) are
described. Section 2.4 quantifies the dependence of assumed distances
on the cloud random velocity. Most of the clouds cannot be made to
belong to the galactic plane population without having equally
improbable peculiar velocities ($> 3\sigma_{V}$; where $\sigma_{V}$
is the cloud-cloud velocity dispersion).

Molecular clouds have been found at high galactic latitudes by
Magnani, Blitz \& Mundy (1985), Keto \& Meyers (1986). Molecular
gas has also been found associated with cirrus at high latitudes by
Des\'{e}rt et al. 1988, Heiles et al. 1989 and Des\'{e}rt et
al. 1990. Clemens and Barvainis (1988) have compiled a catalog of
small clouds in the galactic plane. It would be logical to compare the
properties of the clouds found at high z with the local High Latitude
Clouds (HLCs) and the Giant Molecular Clouds to see if the HACs are
well separated in their properties from the other molecular clouds. In
contrast to the HLCs and the Clemens-Barvainis (CB) objects the HACs
are not selected by optical extinction/infrared emission but by CO
emission. The population of HACs is intermediate in its properties
between GMCs and the local high latitude, high velocity molecular
clouds associated with infrared cirrus and the CB clouds. Part of this
may be due to the different selection criteria. Section 3 describes
the properties of the HACs.

Even though these clouds are at greater distances from the plane than
expected from the scale height of molecular gas, they still are well
within the atomic gas layer, which has about twice the scale height of
molecular gas. The $^{12}CO$ survey by Knapp et al. (1985) from which
the HACs are selected was carried out at the same longitudes as the HI
survey at Arecibo (Bania \& Lockman 1984). We compare the properties
of the molecular gas (as measured by the $^{12}CO(1\rightarrow0)$
line) and HI (in the 21 cm transition) at the locations of
HACs in section 3.

\section{Cloud Identification}
\subsection{The data}

We search for high altitude clouds in the $^{12}CO(1\rightarrow 0)$
survey that Knapp et al. (1985) obtained using the 7 meter antenna at
Bell Laboratories. The data and details of the survey and the
instruments are presented by Knapp et al. (1985).

The survey consists of 38 strip maps (in latitude) of the
$^{12}CO(1\rightarrow 0)$ line emission. The strip maps were taken at
longitudes between $4\deg$ and $90\deg$, spaced at equal intervals in
the $\sin l$, $\Delta \sin l=0.025$. The maps are sampled at intervals
of $\Delta b= 2 \arcmin$ with a half-power beamwidth of $100
\arcsec$. The latitude coverage is $\simeq \pm 2\deg$, varying slightly
from one line of sight to another.  The latitude extent is more than
adequate for studying the tangent point emission (except possibly for
$ l \geq 60\deg$), and extends more than three scale heights both
above and below the centers of distribution in more than half the
lines of sight. The extent of the survey beyond three scale heights at
the tangent points is uneven because the molecular gas layer deviates
significantly from z=0.

The velocity resolution is $0.65 \kms$. The rms noise is typically
0.3 K but varies slightly in each latitude strip. In addition the
$^{13}CO(1\rightarrow0)$ line was mapped with noise levels of 0.1 K
for two longitudes (l=28.36, l=77.16)

We also use the HI survey of Bania \& Lockman (1984) at Arecibo done
at the same longitudes as the CO survey, starting $l \simeq 31$. The
HPBW in the 21 cm line is $4 \arcmin$ and the maps were sampled at
intervals $\Delta b= 2 \arcmin$. The velocity resolution is $1 \kms$.

\subsection{The vertical distribution of the molecular gas}

The vertical distribution of gas at tangent points was modeled for
more than half (23 out of 38) the observed lines of sight.  The low
longitude ($ 4\deg < l <17\deg$) maps were unusable due to the lack of
emission at (or reasonably near) the tangent point velocities. The
high longitudes ($l > 61\deg$) maps were unusable because their
tangent point emission is at low velocities and is contaminated by
emission in reference positions; and because the latitude extent of
the maps is not adequate for this (almost) local emission.

Assuming an axially symmetric model of the inner galaxy, we can obtain
distances to the tangent points, and from the data find the
distribution of the molecular gas in z. We model the distribution as a
Gaussian in {\rm z}, $$\rho(z)=\rho(0) \exp{\left(-\frac{(z-z_0)^2}{2
\sigma_z^2}\right)}$$ fitting both the scale height $\sigma_z$ and the
centroid $z_0$. Examples of best fit models are shown superposed on
the data in Figure 1.

The models are shown as contours of equal temperature in latitude {\rm
l} and velocity {\rm V}. While the {\rm z}-profile is a Gaussian at
each radius, the contours appear to be more complex because the
velocity structure near the terminal velocity must take into account
the velocity dispersion of the gas and the obtrusion of emission from
gas near the tangent points. The details of the modeling are given
in Paper I.

\subsection{Cloud identification}

After fitting the Gaussian distribution in z, we identify clouds that
are more than $2.57 \sigma_z$ (corresponding to a two-sided
significance level of $1\%$) away from the center of the distribution
as High Altitude Clouds (HACs). Since the rms noise is different for
each spectrum, detections are based on the rms noise of each spectrum,
as calculated from the part of the spectrum at velocity $V > (V_T + 3 \
\sigma_V$); where $V_T$ is the terminal velocity and $\sigma_V$ is the
velocity dispersion (determined in paper I). In figure 1, the points
on the b-v diagram denote detections. To be considered at all the
cloud must be detected in more than one adjacent velocity channel and
at more than one adjacent latitude. This imposes a selection effect of
detecting clouds with velocity widths $ > 1.3 \kms$ and a
cross-section in z
$> 9.6 \cos l \pc$. There exists only one cloud more than 2.56
$\sigma_V$ away from the tangent point velocity and that cloud also is
an outlier in z (l=58.21, z=153, V=53.3).

\subsection{Kinematic distances}

In the above discussion the distances have been determined
kinematically, i.e. the gas at extreme velocities is assumed to be at
the tangent point distances (assuming circular symmetry). These
distances could be incorrect due to deviations from orderly circular
rotation of the galaxy. Since we identify high altitude clouds by the
ratio of their distance from the center of distribution $z_0$ to the
scale height of the distribution, these clouds are outliers (but with
different z) if the high and low {\rm z} gas have similar
kinematics. For example if the gas is moving on elliptic rather than
circular orbits, the vertical scale  of a particular tangent point
emission is not correct but the ratio of $z$ to the scale height
$\sigma_z$ is.

These assumptions break down if the kinematics of the HACs are
different from those of the GMCs. High (positive) peculiar velocities
of (relatively) nearby clouds could lead us to misidentify the clouds
as belonging to the tangent point population and overestimate their
distances (and heights above the plane). Figure 2 shows the peculiar
velocity needed by each cloud to be at a smaller z, versus the reduced
height z in units of $\sigma_z$.

Given that the cloud-cloud velocity dispersion is $5-10 \kms$ (Clemens
1985, Stark 1984, Stark \& Brand 1989, Magnani, Blitz \& Mundy 1985,
Malhotra 1994) figure 2 shows that the peculiar velocities needed to
bring many of the HACs close enough to belong to the GMC cloud
population are improbably high.

\section{Properties}
\subsection{Distribution}

Each of the 23 lines of sight analyzed should have two regions (one
above and one below the galactic plane) where one can find HACs,
making 46 regions in all; of these 40 regions are covered in the the
survey of Knapp et al. (1985) and 23 clouds were found, yielding a
probability slightly above a half of finding an outlier to the
population. These numbers however should be interpreted with care as
the clouds are identified as spatially and kinematically distinct
entities on the basis of $3\sigma$ detection levels. There are many
instances of a group of clouds very close in the b-v space and also
lying within the same clump of HI. Any one such group may be a set of
multiple peaks of a faint cloud.

The highest altitude at which a cloud is found is $212 \pc$ (about
four times the average scale height of molecular gas). Of the 23 HACs,
most (18) lie within four scale heights from the center of the
distribution, and two are $\simeq 5.2 \sigma_z$ away. The
z-distribution of HACs is given in figure 3.

\subsection{Properties}

Since this sample of clouds is identified from b-v maps, their
properties (e.g. size) are derived from a single cross-section in z,
i.e. a random chord through the cloud. For a more reliable size and
mass estimate the clouds should be mapped in $l$ and $b$. Most of the
clouds are smaller than GMCs (giant molecular clouds). The smallest
cloud we can detect varies with the longitude. At $l=58.21 \deg$, for
example, a cloud must be $\geq 5\pc$ to be detected in more than one
latitude scan. The largest and the smallest clouds found are 6.8 and
37 $\pc$ in extent (in z). From figure 4 we see that the sizes of the
HACs are distributed in the region between the detection limits and
the typical sizes of GMCs. The size spectrum is consistent with the
size spectrum of GMCs extrapolated to smaller sizes. Figure 4 shows a
histogram of the observed size distribution of the HACs along with the
relative numbers of HACs expected in each size bin if they had the
size spectrum of GMCs ($dn(R)/dR \propto R^{-2}$) (Solomon and Rivolo
1989) normalised to the number of HACs found in the smallest size bin.

The size-linewidth relation seen to be valid for the GMCs (Dame et al
1986)${\sigma_ v} ^2 \propto {R}$
is not valid for the HACs. There exists a very weak
correlation between the size of the cloud and its velocity dispersion
(figure 4) and the slope of the correlation is different; $r \propto
{\sigma_v}^{0.5}$.

The mass is estimated using the standard conversion factor
$X=N(H_2)/I(^{12}CO)=3.0 \times 10^{20} {\rm cm}^{-2} ({\rm K
\kms})^{-1}$ . Considering that we are taking random chords
through a cloud and the expected length of a chord is ($\frac{4}{3}
r$) for a spherical cloud of radius $r$, the mass of the cloud is
given by $M(H_2)=3 \times (R/{\pc}) \times \int {I(^{12}CO) dz dv}
$. The largest mass found is $5.4 \times 10^3 {\msun}$ and the
smallest $ 42 \msun$.

The internal velocity dispersions range from $0.49 \kms$ to $5.8
\kms$. The observed velocity dispersions are higher than
the velocity dispersions derived using the cloud masses, cloud sizes
and the virial theorem. Table 1 gives the ratios of the observed
velocity dispersion and the virial velocity dispersion; this ratio is
significantly higher than 1 for HACs. (It is about 1 for GMCs). Note
that the $\sigma_V(virial)$ depends on $\sqrt{M/R}$ and $(M/R)$ is the
integrated luminosity of the cloud in a b-v map. Therefore
$\sigma_V(virial)$ is a measured quantity. These ratios shows that
either these clouds are not bound by thier own gravity, or they have a
different conversion factor X.

For the line of sight at $l=28\deg$, the HACs are also found to have a
higher ratio of $^{12}CO$ and $^{13}CO$ intensities, ${\it
R}=I(^{12}CO)/I(^{13}CO)$, than the low altitude clouds (cf Polk et
al. 1988). ${\it R}=9.5$ for HACs, higher than the value for galactic
plane clouds; ${\it R}=4$ for the clouds in this sample, and $\simeq
3-6$ for GMCs (Gordon \& Burton 1976; Solomon, Scoville \& Sanders
1979). Mass estimate from $^{13}CO$ observations would make the
discrepancy between the virial mass and the mass inferred from
$^{13}CO$ more pronounced as the velocity dispersions in the two lines
are comparable and the $^{12}CO/^{13}CO$ ratio is larger for HACs.

For comparison the local high latitude cloud population has an average
size of 1.6 {\pc}, average mass $ 40 {\rm M_{\sun}}$ and internal
velocity dispersion between $0.11\ {\rm km s^{-1}}$ and $3.2\ {\rm km
s^{-1}}$ (Magnani et al. 1985). All the high latitude clouds whose
distances are known lie within 3 scale heights of the disk so they are
not outliers in the z-distribution (Blitz 1990).

\subsection{Atomic envelopes}
Atomic and molecular gas show similar large scale distributions and
kinematics in the inner galaxy (e.g. Burton \& Gordon 1976). Comparing
the b-v maps of CO and of HI taken at the same longitudes, both the
galactic plane and the HACs coincide in space (z-distribution) and
approximately in velocity with HI maxima. Figure 6 shows the CO and
the HI emission maps superimposed. The atomic gas associated with the
HACs has greater altitude and velocity extent than the HACs. Table 1
gives the mass of HI associated with a HAC or a group of HACs. In many
cases (e.g. at l=46, figure 6) the HACs have a slightly different
velocity than the peaks of HI surrounding them. The mass of the HI
associated is found to be less than the mass of the molecular gas.

\section{Discussion}
In the previous sections we have established that there exist clouds
making up the higher-than-Gaussian tails of the vertical distribution
of molecular gas. The anomalous heights of most of these clouds cannot
be explained by small discrepancies in kinematics (such as random
motions of clouds). Besides, most of these clouds correspond (in
z-position and velocity) to distinct maxima in HI and are well within
the HI layer.

 The High Altitude Clouds are seen to cover much of the range between
the GMC's and the local high latitude clouds, in their properties. The
requirement that emission be detected in at least two adjacent
latitudes places a lower limit of $\simeq 30 \msun$ on what we would
identify as a cloud. With the number of clouds we have, the size-mass
parameter space seems filled. The size of the clouds varies from
$\simeq 7 \pc$ to $37 \pc$. The mass varies from $\simeq 43 \msun$ to
$5.4 \times 10^3 \msun$. The velocity dispersion covers the range 0.5
and 5.8 $\kms$. Compare this to the local high latitude clouds which
have an average size of 1.6 {\rm pc}, average mass $ 40 \msun$ and
internal velocity dispersion between $0.11 \kms$ and $3.2 \kms$
(Magnani et al. 1985) and the local GMCs which have a typical mass
$1-2 \times 10^5 \msun$, diameter $45 \pc$ and velocity dispersion $2
\kms$.

In spite of their small sizes these clouds do not in general have a
simple velocity structure. Also they are not self-gravitating. Most of
the clouds need to be more massive by about a factor of 25 to be in
virial equilibrium. Again one must take into account the selection of
clouds having velocity widths of $ > 1.3 \kms$, so the small clouds
that are in virial equilibrium are not detected. The present sample of
clouds could be in virial equilibrium if they were elongated in the
direction of the galactic plane with median axis ratios of 25. The
HACs could be virial equilibrium if the $N(H_2)/W(^{12}CO)$ conversion
factor was higher for these clouds. We have calculated the masses
using the $N(H_2)/W(^{12}CO)$ conversion ratio derived from galactic
plane GMCs. It is possible that this factor changes with the height
above the galactic plane. For example this factor is inferred to be
higher for outer galaxy, i.e. the outer galaxy clouds are
underluminous in $^{12}CO$ (Mead \& Kutner 1988). There are two
reasons why we believe the mass derived from CO luminosity rather than
the virial mass. First the metallicity gradient invoked for the high X
value for outer galaxy molecular clouds is perhaps not be justified
for HACs which are a mere 200 $\pc$ away from the galactic plane, and
there is no evidence of metallicity gradient on that scale. Second the
tight correlation between M(CO) and the velocity dispersion that is
seen for the outer galaxy clouds (M(CO) $\propto$ M(virial)) is not
seen here. de Vries et al. (1987) have argued for a lower value of
$N(H_2)/W(^{12}CO)$ for the diffuse molecular gas associated with the
high latitude cirrus on the basis of its infrared emission. The HACs
could thus have a smaller $N(H_2)/W(^{12}CO)$ ratio exacerbating the
discrepancy between M(CO) and M(virial).

While there is atomic gas associated with the clouds, the mass of the
HI associated is typically smaller than the mass of $H_2$. The atomic
gas associated is not enough to gravitationally bind the molecular
clouds. For the one line of sight where $^{13}CO$ was also observed
the ratio of $^{13}CO/^{12}CO$ decreases as we move out of the
galactic plane (Polk et al. 1988), indicating that the high altitude
molecular gas is more diffuse.

Embarrassingly enough one can think of too many explanations for the
presence of high z clouds. Characteristic crossing time for the
vertical movements of clouds through the Galactic plane is $\sim 10^8$
years, whereas the interstellar medium is stirred by supernova
explosions every $\sim 10^7$ years or so. If molecular clouds form in
the Heiles-type shells-supershells (Heiles 1979), it can explain the
presence of HACs. The HACs would then appear along the arcs in l-b
plane corresponding to the supershells. We cannot test this in the
present data set which is sparsely sampled in longitude. For the
galactic plane gas the shells may be difficult to disentangle as the
HACs are still well within the HI layer ($ |l| < 3 \deg$).

The HACs could also be a separate population of clouds with a
different scale height, but in vertical equilibrium in the galaxy by
having a higher cloud-cloud velocity dispersion. It has been suggested
by Stark (1983) that smaller clouds have greater scale heights. HACs
are smaller than the GMCs, so a higher velocity dispersion would go
along in the direction of their having some equipartition of energy
(but not large enough to go as $M^{-0.5}$). With our sample of 23
clouds it is difficult to test if these clouds have a higher velocity
dispersion. However they are coincident in space and velocity with HI
peaks and HI shows an increase in velocity dispersion with height
above the plane (Kulkarni \& Heiles 1987). A separate acceleration
mechanism for small and large clouds could also lead to different
scale heights. There are no characteristic sizes pointing to a bimodal
distribution that one expects of two separate populations. For a
continuously smaller and hotter (higher velocity dispersion)
population of clouds we should see a correlation between the size and
the height above the plane of the HACs. Such a correlation is not seen
for this sample.

Another possibility is that in the HACs we might be witnessing the
formation/dissolution of molecular clouds as they pass from the
galactic plane to less dense regions or vice versa. The lower $H_2/HI$
ratio and $^{13}CO/^{12}CO$ show that the clouds are less dense than
the galactic plane clouds. Lack of virial equilibrium further points
to these clouds being transient objects.

\acknowledgments
It is a pleasure to thank G. R. Knapp for suggesting this problem and
for advice during all stages of this work. We thank F. J. Lockman for
providing the Arecibo HI survey data and R. H. Lupton for his flexible
data reduction and graphics language `SM'. This work was supported by
NSF grant AST89-21700 to Princeton University.

\vfill
\pagebreak

\parindent=0pt
Figure 1: The latitude-velocity contour maps for the longitudes
(indicated at the top right hand corner) where high altitude clouds
are detected. The contour levels are (1.2, 1.6, 4 K).  The best-fit
model of tangent point emission is superimposed (contour levels at
0.2, 0.4, 0.6, and 0.8 times the peak temperature). Points denote
detections at the 3 $\sigma$ level. Broken lines denote 2.57
$\sigma_z$ from the centroid of the z-distribution. A cloud is
identified by detection in more than one adjacent velocity channels
and more than one latitude. A High Altitude Cloud is a cloud found
more than 2.57 $\sigma_z$ away from the local midplane in the tangent
point region (velocities greater than the cutoff velocity indicated by
the vertical line). The blow-ups of the HACs show that many do not
have a simple geometry or a simple velocity structure.

Figure 2: For each cloud in the sample the peculiar velocity needed to
reduce the height estimate is plotted against the reduced height in
units of $\sigma_z$. A (relatively) nearby cloud with a large positive
positive velocity could be mistaken for a tangent point cloud. In such
a case the distance is overestimated so the height above the plane is
also overestimated. In this plot each line represents a High Altitude
Cloud. The overestimate of height is plotted against the velocity
needed to produce that overestimate. A large number of clouds need an
improbably large velocity (more than $3 \sigma_v$, where $\sigma_v$ is
the cloud-cloud velocity dispersion) to have been misidentified as
outliers in z.

Figure 3: The vertical distribution of High Altitude Clouds is shown
in units of (a) $|z/\sigma_z|$ and (b) z. The most outlying clouds are
$\simeq 5.2 \sigma_z$ from the midplane, and the highest cloud is at
$z=212 \pc$.

Figure 4: The size distribution histogram of the High Altitude Clouds
detected. The biggest cloud is $37 \pc$ and the smallest is $7
\pc$. The size-spectrum of Giant Molecular Clouds ($dn(R)/dR \propto
R^{-2}$) normalized to the number of clouds in the smallest bin is
shown by the curve. The High Altitude Clouds have the same size
spectrum as the GMCs.

Figure 5: The sizes (z cross-sections) versus the velocity dispersions
of the High Altitude Clouds. There is a weak size-linewidth
correlation. The solid line shows the best fitting correlation $R
\propto {\sigma_v}^{0.5}$. The dotted line shows the size-linewidth
relation for Giant Molecular clouds $R \propto {\sigma_v}^2$.

Figure 6: The latitude velocity contour maps for HI (light lines) and
CO (dark lines). The contour levels are (10, 20, 30, 40, 50, 60, 70,
80, 90, 100 K for HI) and (1.2, 1.6, 2, 3, 4, 6, 8, 10, 12 K for CO).
There is close correspondence in the positions and velocities of CO
emission with HI peaks both for galactic plane CO as well as for most
of the High Altitude Clouds.

Figure 7: The latitude velocity contour maps of $^{12}CO(1\rightarrow
0)$ (shaded areas) and $^{13}CO(1\rightarrow 0)$ emission (contours:
0.4, 1.2, 4 K) at the longitude $l=28.36 \deg$. There is $^{13}CO$
emission detected for both galactic plane clouds and the high altitude
clouds. The ratio of $^{12}CO$ to $^{13}CO$ however is higher for the
High Altitude Clouds as seen from the side panel which shows $^{12}CO$
(thin lines) and $^{13}CO$ (thick lines) emission integrated over
tangent point velocity range (indicated by arrows).

\makeatletter
\def\jnl@aj{AJ}
\ifx\revtex@jnl\jnl@aj\let\tablebreak=\nl\fi
\makeatother

\begin{planotable}{lrrrrcrr}
\tablewidth{33pc}
\tablecaption{Properties of the clouds}
\tablehead{
\colhead{$l$} & \colhead{z}  &
\colhead{$\Delta z$} &  \colhead{V}    &
\colhead{$\sigma_V$} &   \colhead{$M_{H_2}$}&
\colhead{$\sigma_V$}
& \colhead{$M_{HI}$\tablenotemark{1}}\\[.3ex]
\colhead{$(\deg)$} & \colhead{$(\pc)$}  &
\colhead{$(\pc)$} &  \colhead{$(\kms)$}    &
\colhead{$(\kms)$} &   \colhead{$(\msun)$}&
\colhead{$\overline{\sigma_V(virial)}$}
& \colhead{$(\msun)$}\\[.3ex]}
\startdata
   23.58 &   -148 &   8.9  &   99.4  &   1.63 &  90  & 11.2 & \nodata  \nl
   23.58 &   -212 &   8.9  &   104.4 &   0.69 &  60  &  5.8 & \nodata \nl
   25.25 &    158 &   8.9  &   101.8 &   0.70 &  82  &  5.1 & \nodata \nl
   25.25 &    101 &   8.9  &   105.9 &   0.74 &  99  &  4.8 & \nodata \nl
   28.36 &    74  &  12.9  &   96.7  &   1.92 &  685 &  5.9 & \nodata \nl
   28.36 &    65  &  12.9  &   102.3 &   0.73 & 1109 &  1.7 & \nodata \nl
   30.0  &    80  &  25.4  &   95.1  &   1.93 & 1587 &  5.5 & \nodata \nl
   33.37 &    81  &  36.9  &   99.1  &   2.19 & 4808 &  4.4 & 1096\nl
   33.37 &   -122 &   8.2  &   79.9 &   0.70  &  52  &  6.4 & \nodata \nl
   33.37 &   -197 &  20.5  &   81.2 &   0.74  &  642 &  3.0 & 105\nl
   38.68 &    167 &   7.6  &   81.8 &   1.04  &  123 &  6.2 &  \nodata \nl
   38.68 &    135 &   7.6  &   72.4 &   2.32  & 1193 &  4.4 & 1067\nl
   38.68 &    109 &  15.2  &   74.2 &   0.74  &  215 &  4.7 & 1067\nl
   44.43 &    66  &  35.2  &   59.0 &   1.92  & 3622 &  4.7 & 350\nl
   44.43 &   -195 &  24.6  &   70.5 &   1.04  &  540 &  5.5 & \nodata    \nl
   46.47 &    108 &   6.8  &   48.6 &   0.68  &  43  &  6.8 & 299\nl
   46.47 &    97  &  13.6  &   57.1 &   2.40  & 1665 &  5.5 & 299\nl
   46.47 &   -156 &  34.0  &   57.6 &   2.94  & 3425 &  7.3 & 335\nl
   46.47 &   -201 &   6.8  &   64.4 &   0.67  &  54  &  6.0 & \nodata \nl
   55.59 &    -88 &  24.6  &   46.8 &   5.80  & 5464 &  10.9& 1167\nl
   58.21 &    153 &  12.9  &   53.3 &   1.70  &  956 &  5.7 & 188 \nl
   58.21 &   -84  &   7.8  &   45.2  &   1.72 &  389 &  6.9 & 214\nl
   58.21 &   -136 &   7.8  &   38.7  &   1.00 & 119  &   7.3&  \nodata \nl
\tablerefs{(1) Identical HI masses in two consecutive rows mean that the two
molecular clouds share the same atomic envelope}
\end{planotable}

\vfill
\pagebreak

\end{document}